\documentstyle[aps]{revtex}

\begin{document}
\title{Eavesdropping on the Bostr\"{o}m-Felbinger Communication Protocol in Noisy
Quantum Channel}
\author{Qing-yu Cai}
\address{Wuhan Institute of Physics and Mathematics, The Chinese Academy of Sciences,%
\\
Wuhan, 430071, People's Republic of China}
\maketitle

\begin{abstract}
We show an eavesdropping scheme on Bostr\"{o}m-Felbinger communication
protocol ( called ping-pong protocol ) [Phys. Rev. Lett. 89, 187902 (2002)]
in an ideal quantum channel. A measurement attack can be perfectly used to
eavesdrop Alice's information instead of a most general quantum operation
attack. In a noisy quantum channel, the direct communication is forbidden.
We present a quantum key distribution protocol based on the ping-pong
protocol, which can be used in a low noisy quantum channel. And we give a
weak upper bound on the bit-error ratio that the detection probability $d$
should be lower than $0.11$, which is a requirement criterion when we
utilize the ping-pong protocol in a real communication.
\end{abstract}


\section{\protect\smallskip Introduction}

Quantum key distribution (QKD) is a protocol to be provably secure, by which
private key bit can be created between two parties over a public channel.
The key bits can then be used to implement a classical private key
cryptosystem, such as Vernam cipher [1], some times called a one time pad,
to enable the parties to communicate securely. The basis idea behind QKD is
that Eve can not gain any information from the qubit transmitted from Alice
to Bob without disturbing their states. First, the no-cloning theorem
forbids Eve to perfectly clone Alice's qubit. Secondly, in any attempt to
distinguish between two non-orthogonal quantum states, information gain is
only possible at the expense of introducing disturbance to the signal [1].
Since Bennett and Brassard presented the pioneer QKD work in 1984 [2], there
are a lot of quantum communication protocols today [3-11]. Bostr\"{o}m and
Filbinger presented a deterministic secure direct quantum communication
protocol called ping-pong protocol with a novel security proof [7]. They
show that the security of the ping-pong protocol is unconditional secure
with an abstract mathematics proof. In this letter, we show a scheme that
gives a physical eavesdropping on the ping-pong protocol. Since a real
quantum channel is noisy, we present a modified ping-pong protocol by using
Calderbank-Shor-Steane codes [12] and give a requirement criterion of the
upper bound on the detection probability (bit-error ratio).

The ping-pong protocol utilizes the property that one bit of information can
be encoded in the states $|\psi ^{\pm }>=\frac{1}{\sqrt{2}}(|0>|1>\pm
|1>|0>) $, which are completely unavailable to anyone who has access to only
one of the qubits. To gain information from Alice, Bob prepares two qubit in
the Bell state $|\psi ^{-}>$. Then he stores one qubit (home qubit) and
sends the other one (travel qubit) to Alice through the quantum channel.
Alice can decide the control and the message mode randomly. In the message
mode, Alice performs a unitary operation $\sigma _{z}^{A}$ to encode the
information `1' or dose nothing to encode the information `0'. Then she
sends it back. Bob can get Alice's information by a Bell measurement. In
control mode, Alice performs a measurement in basis $B_{z}=\{|0>,|1>\}$.
Using the public channel, she sends the result to Bob, who then also
switches to the control mode and performs a measurement in the same basis $%
B_{z}$. Bob compares his own result with Alice's result. If both results
coincide, Bob knows that Eve is in line and stops the communication.
Otherwise, Bob sends the next qubit to Alice and the communication continues.

It has been proven that any information Eve gains would make her face a
nonzero detection probability. Eve's aim is to find out which operation
Alice performs. Eve has no access to Bob's home qubit, so all her operations
are restricted to the travel qubit, whose state is completely
indistinguishable from the complete mixture $\rho _{A}=tr_{B}\{|\psi
^{+}><\psi ^{+}|\}=\frac{1}{2}(|0><0|+|1><1|)_{A}$. Bostr\"{o}m and
Felbinger presented an unconditional security proof on the ping-pong
protocol [7]. The most quantum operation is a completely positive map $%
\varepsilon :S(H_{A})\rightarrow S(H_{A})$. One can replace the state of the
travel qubit by the a priori mixture $\rho _{A}=\frac{1}{2}|0><0|+\frac{1}{2}%
|1><1|$, which corresponds to the situation where Bob sends the travel qubit
in either of the states $|0>$ or $|1>$, with equal probability $p=1/2$.
Consider the case where Bob sends $|0>$. Eve adds an ancilla in the state $%
|\chi >$ and perform a unitary operation $E$ on both systems, resulting in 
\begin{equation}
|\psi ^{\prime }>=E|0,\chi >=\alpha |0,\chi _{0}>+\beta |1,\chi _{1}>.
\end{equation}
Defining that 
\begin{equation}
d=|\beta |^{2}=1-|\alpha |^{2},
\end{equation}
obviously, $d$ is the detection probability in the control mode for Eve's
attack. It has been proven that when Alice exactly encodes one bit, the
maximal information Eve gain is equal to the Shannon entropy of a binary
channel, 
\begin{equation}
I_{0}(d)=-d\log _{2}d-(1-d)\log _{2}(1-d).
\end{equation}
The function $I_{0}(d)$ has maximum at $d=1/2$, giving a monotonous function 
$0\leq d(I_{0})\leq 1/2$ on the interval [0,1/2], $I_{0}\in [0,1]$. Any
effective eavesdropping attack can be detected.

\section{a measurement eavesdropping attack strategy}

In their proof [7], Eve's attack operation in line $B\rightarrow A$ is
described as a $completely$ $positive$ $map$ operation, which is an abstract
proof. We will show a practical eavesdropping attack scheme which is
equivalent to a completely positive map attack. Consider that 
\begin{eqnarray}
\sigma _{z}|0 &>&=|0>,\sigma _{z}|1>=-|1>, \\
\sigma _{z}|+ &>&=|->,\sigma _{z}|->=|+>,
\end{eqnarray}
here $|+>=\frac{1}{\sqrt{2}}(|0>+|1>)$, $|+>=\frac{1}{\sqrt{2}}(|0>+|1>)$.
Eve's aim is to find out which operation Alice performs. She can eavesdrop
Alice's information by using the strategy that she attacks the travel qubit
in line $B\rightarrow A$ to prepares the qubit in the state $|\uparrow _{n}>$%
, 
\begin{equation}
|\uparrow _{n}>=\cos \frac{\theta }{2}|0>+\sin \frac{\theta }{2}|1>
\end{equation}
and performs a measurement in the line $A\rightarrow B$ to draw Alice's
information.

Let us analyze the connection between information Eve gained and the
corresponding detection probability she has to face. When Alice sends the
travel qubit to Bob, Eve can capture the travel qubit in the line $%
B\rightarrow A$. she perform a measurement in the basis $B_{z}$. With
probability $p=1/2$, she get $|0>$ or $|1>$. Consider that Eve gets the
result $|0>$. Then the state of the home qubit is immediately collapse to $%
|1>$. Then Eve prepares the travel qubit in state $|\uparrow _{n}>$. In
control mode, the detection probability $d_{m}$ is given 
\begin{equation}
d_{m}=<1|\uparrow _{n}><\uparrow _{n}|1>=\sin ^{2}\frac{\theta }{2}.
\end{equation}
In message mode, the state of the travel is $|\uparrow _{n}>$ after Alice
encoded `0' and becomes 
\begin{equation}
|\downarrow _{n}>=\cos \frac{\theta }{2}|0>-\sin \frac{\theta }{2}|1>,
\end{equation}
when Alice encoded `1'. We calculate the information Eve can gain on this
occasion. Practically, Alice encodes `0' and `1' with equal probability.
After Alice's encoding operation, the state of travel becomes 
\begin{eqnarray}
\rho &=&\frac{1}{2}|\uparrow _{n}><\uparrow _{n}|+\frac{1}{2}|\downarrow
_{n}><\downarrow _{n}|  \nonumber \\
&=&\cos ^{2}\frac{\theta }{2}|0><0|+\sin ^{2}\frac{\theta }{2}|1><1|.
\end{eqnarray}
Then the information Eve can gain from $\rho $ is the classical information
entropy 
\begin{eqnarray}
I &=&-\cos ^{2}\frac{\theta }{2}\log _{2}\cos ^{2}\frac{\theta }{2}-\sin ^{2}%
\frac{\theta }{2}\log _{2}\sin ^{2}\frac{\theta }{2}  \nonumber \\
&=&-d_{m}\log _{2}d_{m}-(1-d_{m})\log _{2}(1-d_{m}).
\end{eqnarray}
Assume that Eve's measurement result is $|1>$ rather than $|0>$. The above
calculation can be done in full analogy, resulting in the same relation.
Thus, one can only use measurement attack in the line $B\rightarrow A$
instead of a $completely$ $postive$ $map$ operation attack. After Eve's such
attack, the entanglement between travel qubit and home qubit does not exist.
Bob can not gain any information from Alice. The information
authentification has to be considered in the line $A\rightarrow B$.

\section{Bostr\"{o}m-Felbinger communication in the noisy quantum channel}

In the ping-pong protocol [7], it only has been considered in the ideal
quantum channel. However, a practical quantum channel is noisy. The noise
can cause the bit errors and the quantum losses. W\'{o}jcik has discussed
the eavesdropping on the ping-pong protocol hiding in the quantum losses
[13]. When Alice and Bob use a noisy quantum channel, they can not
communicate in a direct way, i.e., directly communication must be forbidden.
In control mode, Alice and Bob publish their measurement results. If they
find both results coincide, they stop the communication. In a noisy quantum
channel, a spontaneous disentanglement process is inevitable. Both of the
results will coincide with a nonzero probability. Alice and Bob can not
distinguish whether Eve is in line. The communication has to be stopped.

In a noise quantum channel, direct communication must be forbidden. One can
use the ping-pong protocol to complete a quantum key distribution with CSS
codes [12]. Consider a quantum CSS code Q on $n$ qubits comes from two
binary codes on $n$ bits, $C_{1}$ and $C_{2}$, one contained in the other: 
\begin{equation}
\{0\}\subset C_{2}\subset C_{1}\subset Z_{2}^{n},
\end{equation}
where $Z_{2}^{n}$ is the binary vector space on $n$ bits. Suppose $x\in C_{1}
$ is any codeword in the code $C_{1}$. Then we defined the quantum state $%
|x+C_{2}>$ by 
\begin{equation}
|x+C_{2}>\equiv \frac{1}{\sqrt{C_{2}}}\sum_{y\in C_{2}}|x+y>,
\end{equation}
where $+$ is bitwise addition modulo 2. Suppose that $x^{\prime }$ is an
element of $C_{1}$ such that $x-x^{\prime }\in C_{2}$. Then it has that $%
|x+C_{2}>=|x^{\prime }+C_{2}>$. Hence these codewords correspond to coset of 
$C_{2}$ in $C_{1}$, and this code protects a Hilbert space of dimension $%
2^{\dim C_{1}-\dim C_{2}}$. Since Eve may use a measurement attack in line $%
B\rightarrow A$, the information authentification has to be considered in
line $A\rightarrow B$. The ping-pong protocol can be modified as below. The
modified ping-pong protocol.---(1) Alice creates random $n+m+l$ bits where $m
$ and $l$ is determined by the control parameter $c$. (2) Alice chooses a
random $(n+m+l)$-bit string $b$. She choose a random $v_{k}\in C_{1}$. (3)
Bob creates $m+n+l$ EPR pairs in state $|\psi ^{-}>$. (4) Bob sends half of
each EPR pair to Alice and keeps the others. (5) Alice receives $n+m+l$
qubits. She performs a measurement in basis $B_{z}$ on the $m$ qubit
according to $b$. Alice announces the measurement results and which qubits
she measured.(6) Bob also measurement $m$ qubit. If Bob finds more than $t$
of the measurement results coincide, he aborts the protocol. (7) Alice has
an $n-bit$ string $x$. She performs encoding operation on the $n$ qubit
according to $x$. And does nothing on the $l-$qubit (encoded `0'). And sends
the resulting qubits back to Bob. (8) Bob receives these qubits and performs
a Bell-basis measurements on each EPR pair. (9) Alice announce $b$. If Bob
finds more than $t^{\prime }$ of the $l-$code is `1', he aborts this
protocol. (10) Alice announces $x-v_{k}$. Bob subtracts this from his
result, correcting it with code $C_{1}$ to obtain $v_{k}$. (11) Alice and
Bob compute the coset of $v_{k}+C_{2}$ in $C_{1}$ to obtain the key $k$.
This modified ping-pong protocol not only can protect this communication
against Eve's eavesdropping, but also can protect this communication against
Eve's attack without eavesdropping [14]. This modified ping-pong protocol
can be used in low noisy quantum channel.

The security of a protocol is based on 
\begin{equation}
I(A:B)>\min [I(A:E),I(B:E)],
\end{equation}
where $I(A:B)=H(A)+H(B)-H(AB)$, $H$ is the Shannon entropy [15]. Assume that
the detection is $\sin ^{2}\alpha $, $0\leq \alpha \leq \frac{\pi }{4}$. In
this case, the maximal information Eve can gain is determined by equation
(3), $I(A:E)=I_{0}(\sin ^{2}\alpha )$. We will analyze the maximal
information Bob can gain from Alice when $d=\sin ^{2}\alpha $. The state
Alice and Bob shared after Eve's attack should satisfies the condition $%
d=\sin ^{2}\alpha $. And we want to keeps the entanglement as maximal as
possible. If the entanglement does not exist any longer, Bob can get nothing
from Alice. Although the entanglement is not sufficient, it is necessary in
the communication protocol. According to the qualification described above,
the state may be written in the form as 
\begin{eqnarray}
|\Omega  &>&=\cos \alpha |\psi ^{-}>+\sin \alpha |\phi ^{+}> \\
&=&\frac{1}{\sqrt{2}}(|0>|b_{+}>-|1>|b_{-}>),
\end{eqnarray}
where $|\phi ^{+}>=\frac{1}{\sqrt{2}}(|0>|0>+|1>|1>)$, and $|b_{+}>=\cos
\alpha |1>+\sin \alpha |0>$, $|b_{-}>=\cos \alpha |0>-\sin \alpha |1>$.
Clearly, $|\Omega >$ is a maximal entangled state. When Alice performs an
encoding operation $\sigma _{z}^{A}$, the state $|\Omega >$ becomes $|\Omega
^{\prime }>$%
\begin{equation}
|\Omega ^{\prime }>=\cos \alpha |\psi ^{+}>+\sin \alpha |\phi ^{-}>,
\end{equation}
where $|\phi ^{-}>=\frac{1}{\sqrt{2}}(|0>|0>-|1>|1>)$. Bob performs a
Bell-basis measurement on the EPR pair. With probability $\cos ^{2}\alpha $,
he obtains Alice's codes. With the probability $\sin ^{2}\alpha $, she gains
the wrong codes. Hence the information Bob gained is 
\begin{equation}
I(A:B)=1-[-\cos ^{2}\alpha \log _{2}(\cos ^{2}\alpha )-\sin ^{2}\alpha \log
_{2}(\sin ^{2}\alpha )]\text{.}
\end{equation}
The security of this protocol requires that $I(A:B)>I(A:E)$. In this case,
we can gain the detection probability $d=\sin ^{2}\alpha <0.11$. So we gain
a weak criterion that the detection probability $d$ should be lower than $%
0.11$, $d<0.11$. Since the security of quantum communication should be
unconditional, then the criterion $d<0.11$ should be treated as a
requirement. According to our modified ping-pong protocol, either $t/m$ or $%
t^{\prime }/l$ higher than 0.11, the communication protocol should be
aborted immediately.

\section{conclusion and discussion}

In an ideal quantum channel, a direct communication as a real one-time-pad
key has been discussed a lot [7, 10, 16,17]. In a noisy quantum channel,
direct communication has to be abandoned. Alice and Bob can not distinguish
whether bit-error was caused by noise or caused by Eve. So the communication
has to be stopped.. Fortunately, we can use the direct communication to
complete the quantum key distribution in a noisy quantum channel.

In the ping-pong protocol, for Eve, a measurement attack in line $%
B\rightarrow A$ is equal to a $completely$ $positive$ $map$ operation
attack. But the entanglement between home qubit and travel qubit disappears
under such attack. Bob can not gain any information from Alice. We present a
modified ping-pong protocol which can be used to complete quantum key
distribution in noisy quantum channel. In our modified protocol, security on
both line $A\rightarrow B$ and line $B\rightarrow A$ has been proven. We
give a requirement criterion that the bit-error ratio has to be lower than $%
0.11$, which is an important criterion when we utilize the ping-pong
protocol in a real communication.

\section{references}

[1] M. A. Nielsen and I. L. Chuang, Quantum computation and Quantum
Information (Cambridge University Press, Cambridge, UK, 2000).

[2] C. H. Bennett and G. Brassard, 1984, in $proceedings$ $of$ $the$ $IEEE$ $%
International$ $Conference$ $on$ $Computers$, $Systems$ $and$ $%
\mathop{\rm Si}%
gnal$ $\Pr oces\sin g$, Bangalor, India, (IEEE, New York), pp. 175-179.

[3] A. Ekert, Phys. Rev. Lett. 67, 661 (1991).

[4] D. Bru\ss , Phys. Rev. Lett. 81, 3018 (1998).

[5] H. Lo and H. F. Chau, Science 283, 2050 (1999).

[6] A. Beige, B.-G. Englert, C. Kurtsiefer, and H. Weinfurter, Acta Phys.
Pol. A 101, 357 (2002).

[7] K. Bostr\"{o}m and T. Felbinger, Phys. Rev. Lett. 89, 187902 (2002)

[8] C. H. Bennett, Phys. Rev. Lett. 69, 2881 (1992).

[9] G. Brassard, N. Lutkenhaus, T. Mor and B. C. Sanders, Phys. Rev. Lett.
85, 1330 (2000).

[10] Q.-y. Cai and B.-w. Li, Chin. Phys. Lett. 21(4),601 (2004).

[11] V. Scarani and N. Gisin, Phys. Rev. Lett. 87, 117901 (2001).

[12] Calderbank A R and Shor P W 1996 Phys. Rev. A 54 1098; Steane A M 1996
Proc. R. Soc. London A 425 2551.

[13] A. W\'{o}jcik, Phys. Rev. Lett. 90, 157901 (2003).

[14] Q.-y. Cai, Phys. Rev. Lett. 91, 109801 (2003).

[15] I. Csisz\'{a}r and J. K\"{o}rner, IEEE Trans. Inf. Theory 24, 339
(1978).

[16] F.-G. Deng, G.-L. Long and X.-S. Liu, Phys. Rev. A 58, 042317 (2003).

[17] Q.-y. Cai and B.-w. Li, Phys, Rev. A (to be published).

\end{document}